\documentclass[pre,a4paper,oneside,twocolumn,nofootinbib]{revtex4}

\usepackage[english]{babel}
\usepackage[latin1]{inputenc}
\usepackage{graphicx}  
\usepackage{latexsym} 
\usepackage{amssymb,amsmath,amsthm,amsfonts}  
\usepackage{color} 
\usepackage{epstopdf}

\usepackage{hyperref} 
\usepackage{attachfile}

\usepackage[normalem]{ulem} % Pacote para comando de riscar texto: \sout{texto}

\begin{document}

\title{The metabolic origins of big size in aquatic mammals}

% Titulos antigos:

% The metabolic origins to explain why some aquatic mammals are so big}\\

% \textit{ Why are aquatic mammals big?}}

%\title{Metabolic Scaling Theory explain the Gigantism Effect and the relations between body fat and longevity in marine mammals: the metabolic-catabolic minimisation hypothesis}

\author{
William Roberto Luiz S. Pereira\textsuperscript{1}\textsuperscript{$\diamond$}
and Fabiano L. Ribeiro\textsuperscript{2}\textsuperscript{*}}
%\affiliation{.}

\date{\today}

\begin{abstract}
The group of large aquatic mammals has representatives being the largest living beings on earth, surpassing the weight and size of dinosaurs. In this paper, we present some empirical evidence and a mathematical model to argue that fat accumulation in marine mammals triggers a series of metabolic events that result in these animals' increased size. Our study starts by analysing 43 ontogenetic trajectories of species of different types and sizes. For instance, the analyses include organisms with asymptotic mass from 27g (Taiwan field mouse) to
$2.10^{7}$g (grey whale). The available data allows us to determine all available species' ontogenetic parameters (catabolism and anabolism constant, scaling exponent and asymptotic mass).
The analyses of those data show a  minimisation of catabolism and scaling exponent in marine mammals compared to other species analysed. We present a possible explanation for this, arguing that the large proportion of adipose tissue in these animals can cause this minimisation. 
That is because adipocytes have different scaling properties in comparison to non-adipose (typical) cells, expressed in reduced energetic demand and lower metabolism.
The conclusion is that when we have an animal with a relatively large amount of adipose tissue, as is the case of aquatic mammals,  the cellular metabolic rate decreases compared to other animals with the same mass but with proportionally smaller fat tissue. 
A final consequence of this cause-effect process is the increase of the asymptotic mass of these mammals.
%Our conclusion is based on the empirical evidence that typical cells and fat cell (adipocytes ),have different scaling properties,   which affect the ontogenetic parameters of this animals. 
%on empirical evidence collected from the literature, composed of 
\end{abstract}

\maketitle

\textbf{1} Independent Researcher.

\textbf{2} Departmento de Fisica (DFI), Universidade Federal de Lavras (UFLA), Lavras MG, Brazil;

%\textbf{*} Correspondence: fribeiro@ufla.br

\textbf{*} fribeiro@ufla.br

\textbf{$\diamond$} william.roberto.luiz@gmail.com

%\textbf{1} Department of Physics (DFI), Federal University of Lavras (UFLA), Lavras MG, Brazil;


\section{Introduction}\label{introduction}

The blue whale (\textit{Balaenoptera musculus}) is the largest and heaviest living being, having up to 180 tons \cite{Bianucci2019}, \cite{Vermeij2016}. It is  30 times bigger than the heaviest land animal, the African elephant \textit{(Loxodonta africana)}, weighing around 6 tons \cite{HANKS1972}\cite{Schiffmann2020}, and twice the size of the largest terrestrial animals that ever lived, the argentinosaurus (\textit{Argentinosaurus huinculensis}) \cite{Benson2014}, a species of Sauropoda dinosaur. 
%(see Fig.~(\ref{fig_animais}) for comparison).
This enormous difference in size between whales and terrestrial animals has to do, in part, with the reduced gravity effect in the aquatic environment
%, because it's given that gravity limits growth in terrestrial environments 
\cite{Sander2011}. 
However, what other factors, complementary to gravity, are acting on the appearance and persistence of such large marine animals?

%The size of the blue whale is so impressive that it is twice the size of the largest terrestrial animals lived ever, such as the Sauropoda (\textit{Dreadnoughtus} sp.), a kind of dinosaur (see Fig.~(\ref{fig_animais}) for comparison). 

% Some default explanations to the appearance of large animals -- the so-called  

Some common explanations for the appearance of large animals -- the so-called  \textit{gigantism effect} \cite{Goldbogen2018} -- is that essential resources need to be abundant and effectively recycled and reused in a highly developed ecological infrastructure; this is a rare biological condition \cite{Vermeij2016}. Moreover,  large animals have an advantage against predators \cite{Ford} and also improve the capacity to forage food \cite{Goldbogen2018}. 
Nevertheless, there are also other non-trivial explanations for the phenomenon, such as genetic adaptations in the transition from land to water lifestyle \cite{Huelsmann2019}, intense environmental pressure (the Contingency Rule (\cite{Turchin2002}), and the evolutionary memory to favour biomass accumulation\footnote{The biggest aquatic animals, which include the Pinnipedia (seals) and Cetaceans (whales), have a common ancestor with the common hippo \textit{(Hippopotamus amphibius)} \cite{Maust-Mohl2019}, one of the heaviest land animals (up to 3 tons). It suggests that marine mammals have a genetic framework and an evolutionary memory to favour biomass accumulation.}.
To sum up, the gigantism effect is, in fact, a multifaceted phenomenon in which each factor cited above, among others, contributes to some degree (see \cite{Goldbogen2018}).

%But there are also other non-trivial explanations for that, as genetic adaptations in the transition from land to water lifestyle \cite{Huelsmann2019}, strong environmental pressure  (the Contingency Rule (\cite{Turchin2002}), and the evolutionary memory to favour biomass accumulation\footnote{The biggest aquatic animals, which include the Pinnipedia (seals) and Cetaceans (whales), have a common ancestor with the common hippo \textit{(Hippopotamus amphibius)} \cite{Maust-Mohl2019}, one of the heaviest land animals (up to 3t). It suggests that marine mammals have a genetic framework and an evolutionary memory to favour biomass accumulation.}, just to name a few. It suggests that marine mammals have a genetic framework and an evolutionary memory to favour biomass accumulation.}, just to name a few.

One trademark of marine mammals is their capacity to stock fat, especially in pinnipeds (seals), sirenia (manatees) and cetaceans (whales). The thickness of the fat layer (blubber) in cetaceans reaches 20 cm (for example, in \textit{Eschrichlius robustus}) and makes up from 15 to 55\% of the body mass \cite{Ryg1993,Wang2015,Kershaw2018,Pond1978}. 
In addition to energy storage, the blubber acquired many physiological and physical functions, such as thermal insulation, aid in flotation and locomotion, and increasing swimming efficiency by smoothing the body contour \cite{Wang2015}.

%around 30\% of the total body mass (for example, in \textit{Balaenoptera physalus}) \cite{Pond1978}. 

Empirical evidence suggests that the way to store fat evolved from single-celled organisms (bacteria and yeasts) to specialized multi-cellular beings.  While single-celled organisms developed regions accumulating lipid in droplets, the multi-cellular organisms developed their own adipose organ (in fish, amphibians, and reptiles) with the subsequent organization of subcutaneous adipose tissue (in mammals) \cite{Birsoy2013}.
It shows that the \textit{adipocytes}, cells that store fat and compose the adipose tissue,  had their very evolutionary history, which occurred along with the evolution of large taxonomic groups. 
%As the adipocyte assumed cellular identity in the metazoans, it acquired particularities initial aimed at storing energy, but its plasticity favoured the attribution of new physiological functions.It is reflected in morphological, biochemical and genetic aspects, highly observed in marine mammals.
%\textcolor{red}{this paragraph seems too specialized for me -- can it be rephrased so that non-experts understand it?}

%\textit{adipocytes}, are also called \textit{adipose cells}, 

%Adipocytes, sao celulas que armazenam gordura e compoem o chamado adipose tissue. Estas celulas tambem sao conhecidas por lipocytes or fat cells, and in some case it volume is composed of 80\% of fat.

%In this work we try to connect this fat properties with the metabolism and the ontogenetic 

Here we connect these fat properties with the metabolism and the ontogenetic scaling properties, specially using as a starting point  the theories developed by West et al.  \cite{West1997,West2001,West2002,West2004} and expanded by other researchers \cite{Banavar2002,Savage2007,Moses2008,Sibly2015,DeLong2010,ribeiro_tumor2017}.
Those theories try to explain the empirical evidence  that the \textit{metabolic rate} $R$ of an organism obeys an allometric scaling law with its mass $m$ in the form \cite{Kleiber1975}

\begin{equation}\label{eq_Kleiber's Law}
R = R_0 m^ {\beta} \, . 
\end{equation}
This relation is known by Kleiber's law, where $R_0$ is the \textit{allometric constant} and $\beta$ the \textit{scaling exponent}. 
Empirical evidence suggests that $\beta<1$, which implies that larger animals are more efficient energetically -- demanding less energy per cell \cite{West2001}.
\textit{Kleiber's Law} is valid in inter-species context (i.e.\ using adult mass of different species) and in intra-species context during \textit{ontogenetic process} (using time evolution of mass of a single species) \cite{Moses2008,Zuo2009}.

%The relation~(\ref{eq_Kleiber's Law}), namely \textit{Kleiber's Law},  is valid in  inter-species context (i.e. using adult mass of different species)  and   in intra-species context during \textit{ontogenetic process} (using time evolution of mass of a single species) \cite{Moses2008,Zuo2009}. 

West et al. \cite{West1997} explain such scaling properties as a transport optimisation process. 
Natural selection operates in the efficiency of resource distribution, generating an optimum network distribution where the calibre of the vessels is hierarchically decreased until capillaries at the lowest level of branching that are invariant.
%until an invariance at the lowest level of branching (capillaries).
This optimum network distribution reduces energy expenditure on transportation, leading to an optimal value of $\beta=3/4$ in vascular multi-cellular Metazoa \cite{DeLong2010}. The original West et al. model is derived with details in \cite{Ribeiro2022}.
%This optimum network distribution reduces energy expenditure on transportation, reflecting in an optimal value of $\beta=3/4$ in vascular multi-cellular Metazoa \cite{DeLong2010}.

All hypotheses and models to explain transport optimization were intensively debated and improved \cite{Glazier2014} since the the publication of West et al. work in the later 90s \cite{West1997}.
%All hypotheses and models to explain transport optimization were intensively debated and improved \cite{Glazier2014} since the the publication of the original work of West et al., in later 90s \cite{West1997}.
However, there is no theoretical background yet to explain why some organisms deviate to lower values of $\beta$ from the expected values predicted by the West et al. theory.
For instance, there are mathematical foundations to explain the superior-limit $\beta \to 1$ from microscopic interactions between non-specialized cells \cite{ribeiro_tumor2017},  
but to our best knowledge lower $\beta$ values have not been treated from the metabolic/scaling point-of-view.
%but nothing in the literature, unless for our knowledge, discuss lower $\beta$ values from the metabolic/scaling point-of-view.

There are substantial empirical metabolic scaling findings in virtually all taxonomic groups and in various experimental conditions/designs. In relation to theoretical studies, Glazier \cite{Glazier2014} describes four research lines to explain biological scaling: (I) surface-area hypothesis, (II) network of resource distribution hypothesis, (III) system composition hypothesis, and (IV) resource demand hypothesis. This classification helps us to organize the hypothesis, theories and experimental designs, even though these theories are not exclusionary, mainly III and IV, where our work is based.

In the present work, we try to shed some light on this discussion by combining theoretical (analytical) and experimental data, showing and explaining the small values of the metabolic scaling exponent that we observed in marine mammals (details in section~(\ref{sec_empirical})).
More specifically, we use a careful methodology to fit curve-to-data from ontogenetic trajectories to show that aquatic mammals present a scaling exponent $\beta$ significantly smaller than 3/4 -- the value predicts by West et al. theory. 
We justified this trend and also the increased size in these animals by the large composition of their adipocytes. 
More specifically, we offer some empirical findings and a theoretical approach to argue that {\bf fat accumulation} in aquatic mammals triggers a series of events that culminates in the increase in the size of these animals.

%we present theoretical and empirical evidence suggesting that the big size of marine mammals is a response to their fat accumulation. 

%We justified this trend and also the increased size in these animals as triggered by the large composition of their adipocytes. 

%The paper is organized in the following way. 
The paper is organized as follows.
In Section~(\ref{sec_model}), we present the ontogenetic growth model -- and its parameters -- on which our analyses will be based.
In section~(\ref{sec_empirical}), we present our analyses for 43 species, in which we get the ontogenetic parameters for each species. 
In section~(\ref{sec_catabolism_fat}), we discuss the role played by the fat tissue in marine mammals and present a mathematical model to demonstrate how the scaling properties of adipose cells yield a minimization of the energetic demand in such animals. 
%In section~(\ref{sec_catabolism_fat}), we argue the play ruled by the fat tissue in marine mammals and present a mathematical model to demonstrate how the scaling properties of adipose cells yield to a minimization of the energetic demand in such animals. 
In section~(\ref{sec_biological}) we present some biological foundation to the arguments proposed. 
The paper finish with final considerations in section~(\ref{sec_final}).

%\begin{figure}
%	\begin{center}
%	\includegraphics[width=\columnwidth]
%	{Animals_comparison.png}
	%{Dreadnoughtus-size-fossil-sauropod-2009.jpg}
	%\includegraphics[width=\columnwidth]{Dr-kMb4U4AAMcWv.jpg}
%		\caption{ \label{fig_animais} 
%		Comparative image showing the magnitude of different species. 	The size of the blue whale is so impressive that it is twice the size of the largest terrestrial animals lived ever, such as species of Sauropoda group (for example, \textit{Argentinosaurus huinculensis}), a kind of dinosaur.
%  \textcolor{red}{(source of Figure?)}
 %  https://pixabay.com/pt/vectors/akimbo-macho-homem-silhueta-2026939/
 %
 %https://publicdomainvectors.org/pt/vetorial-gratis/Ilustra%C3%A7%C3%A3o-em-vetor-silhueta-de-baleia/14207.html
  
%		}
%	\end{center}
%\end{figure}

\section{Ontogenetic growth model}
\label{sec_model}

The ontogenetic models, from Bertalanffy and Richards's primordial works \cite{bertalanffy1957,richards1959} to the most advanced and contemporary studies \cite{West2001,Banavar2002, Ribeiro2017}, have been successful in describing individual organism growth.
Specifically in the seminal work of West et al. \cite{West2001}, the authors derive the logistic shape of the temporal organism growth considering that the total energy metabolised can be used either to create new cells -- the \textit{anabolism} -- or to maintain existing cells -- the \textit{catabolism}.

The idea can be expressed in the mathematical form
\begin{equation}\label{eq_edoR}
R =  E_c \frac{dN}{dt} + N R_c \, , 
\end{equation}
where $R$ is the \textit{metabolic rate} (measured in Watts, i.e. Joules per second). 
The first term on the right of this equation, $E_c dN/dt$, is the energy per time $dt$ spend to create $dN$ new cells, with $E_c$ being the energy necessary to create one new cell, also called \textit{activation energy} \cite{West2004}\footnote{The activation energy $E_c$ is a meaningful but also controversial variable. There is a paucity of empirical observations and any well-established experimental design to measure it.}. 
The second term on the right of Eq.~(\ref{eq_edoR}), $N R_c$, is the energy per time $dt$ to maintaining the existing $N$ cells, and $R_c$ is the \textit{cellular metabolic rate}, i.e. the energy necessary to maintain one cell.  

The metabolic rate also obeys the allometric law Eq.~(\ref{eq_Kleiber's Law}), and then if  $m_c$ is the mass of a single cell and $m = N m_c$ is the mass of the organism, then Eq.~(\ref{eq_edoR}) leads to
\begin{equation} \label{edo_m}
\frac{d m}{dt} = A m^{\beta} - B m \, .
\end{equation}
Here, we introduce 
\begin{equation}\label{Eq_A}
A \equiv \left( \frac{R_0 m_c}{E_c} \right)\, ,
\end{equation}
namely the \textit{anabolism constant} (measured in $\textrm{grama}^{1-\beta}/\textrm{time}$), which defines the rate at which the mass is incorporated into the organism. 
Note that $A\propto 1/E_c$, which means that this parameter is a measure of the energy necessary to create a new cell.
We also introduce 
\begin{equation}\label{Eq_B}
B \equiv  \frac{R_c}{E_c}\, , 
\end{equation}
namely the \textit{catabolism constant} (measured in unit of frequency $1/\textrm{time}$), which defines the rate at which the organism uses the energy for vital demands. Eq.~(\ref{Eq_B}) expresses that the catabolism can also be understood as the relation between cellular metabolic rate and activation energy. 
%That means that when B is small (the situation for aquatic mammals, conform will be presented in the next section) then celular energy is very smaller than the energy to create a new cell...

%In addition, $B$ measures the ratio between the cellular metabolic rate and the activation energy. 

%Equalling $E_c$ in Eqs. (\ref{Eq_B}) and (\ref{Eq_A}) yields to write the cellular metabolic rate as 

Eliminating $E_c$ in Eqs. (\ref{Eq_B}) and (\ref{Eq_A}) yields
\begin{equation}\label{eq_RcAB}
R_c \propto \frac{B}{A} \, ,   %A \cdot B.    
\end{equation}
that means cellular metabolic rate can be inferred by the anabolism-catabolism relation. 
The anabolism constant $A$ is invariant among species of the same taxonomy group, 
%since in this case it depends only on scaling invariant parameters\footnote{However $A$ must to differ from distinct taxonomic groups, since it depends directly on the alometric constant $R_0$,  which varies between different taxonomic groups.}.  
since it depends only on scaling invariant parameters\footnote{However, $A$ must to differ between distinct taxonomic groups, since it depends directly on the allometric constant $R_0$, which varies between taxonomic groups.}.
This fact, together with Eq.~(\ref{eq_RcAB}), suggest that inside the same taxonomy group the cellular metabolic rate can be understood solely by the catabolism constant, i.e.\ $R_c \propto B$.

% Using some experimental data, as  $R_0 = 1,9.10^{-2}$ watts$/g^{\beta}$; $m_c \approx 3.10^{-9}$g; e $E_c \approx 2,1 . 10^{-5}$J \cite{west???} and considering $\beta = 3/4$ ???, \textcolor{red}{ver as referencias que o west cita ... ....} it is possible get a empirical value for the anabolism constant as 

% \begin{equation}
%  A \approx 0,234 \frac{\textrm{g}^{\frac{1}{4}}  }{\textrm{day}}.
% \end{equation}

%Coming back to Eq.~(\ref{edo_m}), it has as solution

Equation~(\ref{edo_m}), in turn, has as solution
\begin{equation}\label{solucao_edo}
m(t) = \left[ \frac{A}{B} + \left(m_0^{1- \beta} - \frac{A}{B}\right)e^{B(\beta -1)t}   \right]^{\frac{1}{1-\beta}} \, ,
\end{equation}
where $m_0$ is the initial mass of the organism.
This solution diverges for $t$ sufficiently large when $\beta>1$ and $B>0$, or when $\beta<1$ and $B<0$.
However, for $\beta <1$ (sublinear regime) and $B>0$, 
%which are compatible with biological systems\cite{West2001}, 
which is in agreement with biological systems\cite{West2001}, this solution converges to  
\begin{equation}\label{eq_M}
m(t \gg 1) \equiv M = \left (\frac {A} {B} \right)^{\frac{1}{1- \beta}} \, .
\end{equation}
Here we define $M$ as the mature (asymptotic) mass of the organism. 
For the special case that $\beta<1$ and $B~\approx~0^+$, and according to the Eq.~(\ref{solucao_edo}), the mass growth initially as power law (given by $m(t)\sim t^{\frac{1}{1-\beta}}$) and then saturates to $M$ \cite{ribeiro_tumor2017}.
More details about the solution of this model are presented in Appendix~\ref{app_ontogenetic}. 

%With this sublinearity of $\beta$, it is possible to write the solution for Eq.~(\ref{edo_m}) in terms of $A$ (the classical Bertalanffy-Richards solution) or  in terms of  $B$ (West-Banavar\cite{???} solution). See Appendix~(\ref{???}) for more details.

\section{Empirical Analyses}\label{sec_empirical}

%The empirical investigation used in this work sums 43 ontogenetic trajectories collect from the literature, including organisms with asymptotic mass from 27g (Taiwan field mouse) to $2.10^{7}$g (grey whale).
%The available data and the methodology described in  Supplemental Materials~(\ref{append_metod}) and~(\ref{append_planilhas}) allows us to determine the ontogenetic (macroscopic) parameters ($A$, $B$, $\beta$ and $M$) for all available species.
The empirical base used in this work consists 43 ontogenetic trajectories collect from the literature, including organisms with asymptotic mass ranging from 27g (Taiwan field mouse) to $2\cdot 10^{7}$g (grey whale).
The data and the methodology, described in Supplemental Materials~(\ref{append_metod}) and~(\ref{append_planilhas}), allow us to determine the ontogenetic (macroscopic) parameters ($A$, $B$, $\beta$ and $M$) for all included species.

%20368000g  whale 
%2.3t (elephant seal).
%(Taiwanese wood mouse, Apodemus semotus) to 2.3t (male elephant seal, Mirounga leonina).

\begin{figure}
	\begin{center}
	\includegraphics[width=\columnwidth]{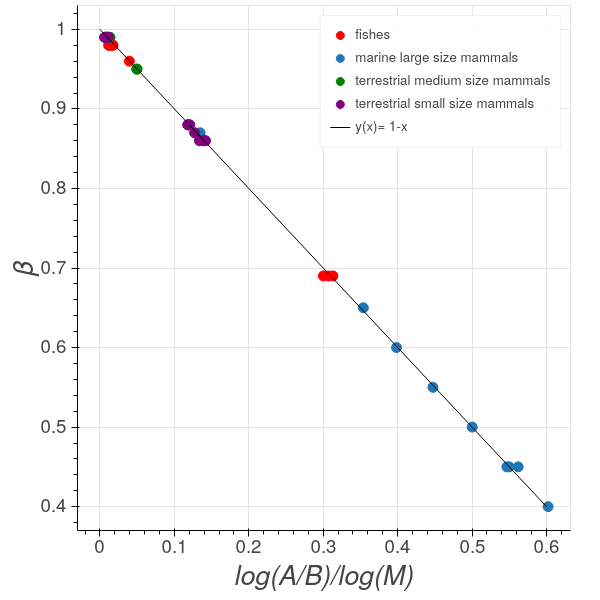}
	\caption{ \label{fig_ABMbeta} 
	Graph showing how the four ontogenetic parameters ($A$, $B$, $\beta$ and $M$) are interconnected one each other. The straight line,  given by Eq.~(\ref{eq_M}), fits the data very well, regardless of the species.
	It means that any change in one of these parameters  reverberates automatically in the other parameters, always maintaining the constraints imposed by Eq.~(\ref{eq_M}).}
	\end{center}
\end{figure}

%The first result that we can identify is the strong inter-correlation between the numeric values assumed by these four ontogenetic parameters, as suggested by the results present in Fig.~(\ref{fig_ABMbeta}). 
%Over there, one can see that all analysed species, independently of their type or size,  obey the relation~(\ref{eq_M}) rigorously. 
%We can infer by this result that any change in one of these parameters - caused by adaptation, for instance - reverberates automatically in the other parameters, always maintaining the constraints imposed by Eq.~(\ref{eq_M}).
The first finding that we can obtain after estimating the four ontogenetic parameters are the strong correlations between their numeric values, as suggested by the results presented in Fig.~(\ref{fig_ABMbeta}). 
One can see that all analysed species, independently of their type or size, rigorously obey Eq.~(\ref{eq_M}). 
From this result we can infer that any change in one of these parameters -- for instance caused by adaptation -- automatically reverberates in the other parameters, always maintaining the constraints imposed by Eq.~(\ref{eq_M}).

\begin{figure}
	\begin{center}
	\includegraphics[width=\columnwidth]{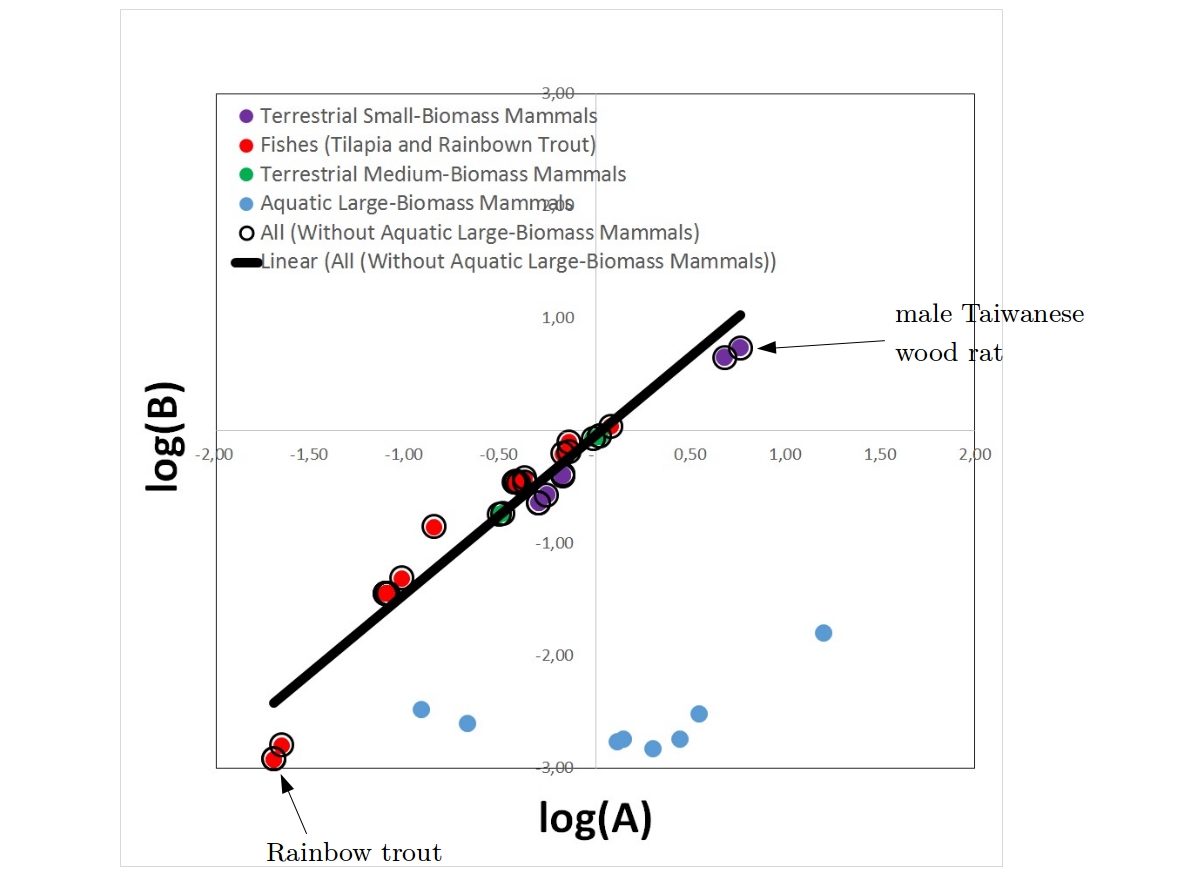}
	\caption{ \label{fig_BxA} 
	The catabolism constant $B$ as a function of the anabolism constant $A$ (in a log-log plot). The data suggest a strong linear relationship between these constants,
	and the slope is related to the cellular metabolic rate ($R_c$), conform Eq.~(\ref{eq_RcAB}). 
However the marine mammals (blue points)
break this rule, presenting a much lower catabolism ($B$) than should be (according to this rule).  %(linear relation between $A$ and $B$). 
The rainbow trout was the animal with the lowest anabolism and catabolism constants among the analysed species, while the male Taiwanese wood rat is the species with the highest numerical value for these metabolic constants.
}
\end{center}
\end{figure}

%However, when it is analysed sub-groups of these parameters, some particularities can be observed, especially what concerns marine mammals.  For instance, there is an evident differentiation between the marine mammals and the other analysed species in respect to the anabolism-catabolism relation.
However, when sub-groups of ontogenetic parameters are analysed, some particularities can be observed, especially concerning marine mammals.
For instance, there is an evident differentiation between the marine mammals and the other analysed species in respect to the anabolism-catabolism relation.
One can see in Fig~(\ref{fig_BxA}) that there is a strong linear dependence between $A$ and $B$ (and the slope is related to $R_c$, cf.\ Eq.~(\ref{eq_RcAB})), which implies that animals with higher anabolism also tend to have higher catabolism.
However, marine mammals break this rule, maintaining their catabolism ($B$) much lower than it should be (according to this linear rule).
This result, together with  Eqs.~(\ref{Eq_B}) and~(\ref{eq_RcAB}), suggests that these mammals have lower cellular metabolic rates. 

\begin{figure}
	\begin{center}
\includegraphics[width=\columnwidth]{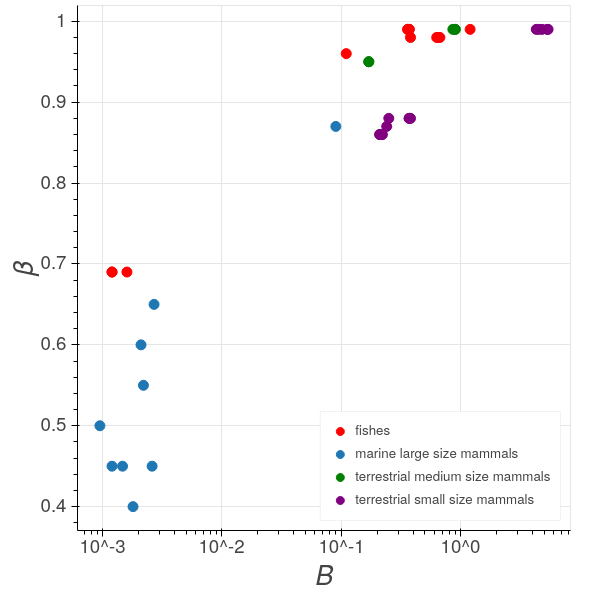}
		\end{center}
	\caption{ \label{fig_beta-x-B} 
	Linear-log graph of the scaling exponent $\beta$ and the catabolism constant $B$.
	%Once again one can see that 
	The marine mammals (blue points) differentiate from the other species: they are the biggest animals between the analysed species and, at the same time, the ones with lower $\beta$  and $B$ values. It suggests that these large mammals' metabolism is relatively slower compared to the other analysed species. }
\end{figure}

The catabolism reduction in marine mammals is also accompanied by a lower scaling exponent $\beta$ among these animals. 
The scaling exponent of the considered terrestrial mammals lies in the interval $0.70 <\beta < 0.99$.
%, whose variability is following what we expected by the theory; 
However, the scaling exponent for the analysed marine mammals is $\beta \approx 0.5$, that is much lower than the other species.
Fig~(\ref{fig_beta-x-B}) shows that these large mammals differ from the other species in terms of both $\beta$ and $B$. % values.
%, presenting lower values for theses parameters.   
Marine mammals are the biggest animals among the analysed species and, at the same time, the ones with lowest scaling exponent ($\beta$) and catabolism constant ($B$). 
Consequently, this large mammals' metabolism is relatively slow compared to the other analysed species.
%These facts suggest that this large mammals' metabolism is relatively slow compared to the other analysed species.
It is valid to call the attention that the minimisation of these parameters in marine mammals is constrained by the relation~(\ref{eq_M}), which apparently governs the connection between the ontogenetic parameters ($A$, $B$, $M$ and $\beta$). 

% paragraph seems redundant:
%The result presented in this section allows us to conclude the following. 
%Marine mammals present relatively lower metabolism since they have lower scaling exponent and catabolism constant. However, the minimisation of these parameters is done constrained to the relation~(\ref{eq_M}), which apparently governs the connection between the ontogenetic parameters ($A$, $B$, $M$ and $\beta$). 

%\textcolor{red}{Falar alguma coisa aqui sobre a relacao entre beta e A/B, dado pela forma (from   eq~\ref{eq_M}): }

%\begin{equation}
%\beta = 1- \frac{1}{M} %\ln\left(\frac{A}{B}\right)    
%\end{equation}
%\textcolor{red}{If $A=B$ then $\beta =1$;  if $B \to 0$ then $\beta$ is minimized...}

%\section*{Comments about extreme cases}
%To finalise this section, we would like to discuss the two species of our database that present the extreme values of anabolism and catabolism constant: the rainbow trout and the Taiwan field mouse. 
%By these examples, the intention is showing that the organism's energetic demand -  its metabolism -  reflects directly in the values of $A$, $B$ and $\beta$.

\section*{Extreme cases}

To finalise this section, we would like to discuss two species of our database that exhibit extreme values of the anabolism and catabolism constants, i.e.\ the rainbow trout and the Taiwan field mouse. 
The intention is to show that the organism's energetic demand -- its metabolism -- reflects directly in the values of $A$, $B$ and $\beta$.

The rainbow trout (\textit{Oncorrhynchus mykiss}), which lives in the Pongokepuk River, a low-temperature region in Alaska \cite{MacDonald_Lisac_1996}, is the animal with the lowest anabolism and catabolism constants among the analysed species (see Fig.~(\ref{fig_BxA})). 
It is valid to say that ectothermic animals, like this one, have metabolism directly influenced by temperature \cite{Audzijonyte2020}. 
%That means, the slower metabolism of this fish (small $B$ and $A$, simultaneously) can be explained as a result of the lower temperatures environment it lives.  
That means that the slower metabolism of this fish (small $B$ and $A$, simultaneously) is associated with the lower temperatures in its environment. Moreover, this species exhibits a smaller scaling exponent than the other non-marine mammal's species ($\beta \approx 0.69$).
%Maybe as a consequence of this, this species present smaller scaling exponent than the other non-marine mammal's species ($\beta \approx 0.69$).

In contrast, the male Taiwan field mouse (\textit{Apodemus semotus}) is the species with the highest value of the anabolism and catabolism constants (see Fig.~(\ref{fig_BxA})), which is evidence of high metabolism.
In fact, this species reaches sexual maturity very quickly (in 25 days after being born) \cite{Lin1993}, expressing a high growth rate (related to anabolism), demanding a high metabolism. 
Moreover, this animal exhibits a larger value of the scaling exponent ($\beta = 0.99$), which is also compatible with this rodent's high metabolic demand. 

%The analysis of these extreme cases reveals to be consistent with the assumption that the energetic demand of an organism reflects directly on the ontogenetic parameters. 
%That is, species with \textit{lower metabolism} also have smaller anabolism, catabolism and scaling exponent. 
%Apparently, that is also happening in marine mammals. 

These extreme cases reveal to be consistent with the assumption that the energetic demand of an organism is reflected in the values of the ontogenetic parameters. 
Species with \textit{lower metabolism} also have smaller anabolism, catabolism and scaling exponent. 
%Apparently, that is also happening in marine mammals.
As we will see below, this also holds true for marine mammals.

%%%%%%%%%%%%%%%%%%%%%%%%%%%%%%%%%%%%%%%%%%%%%%
\section{Metabolism and the fat stores in marine mammals}\label{sec_catabolism_fat}
%%%%%%%%%%%%%%%%%%%%%%%%%%%%%%%%%%%%%%%%%%%%%

\begin{figure}
	\begin{center}
			\includegraphics[width=\columnwidth]{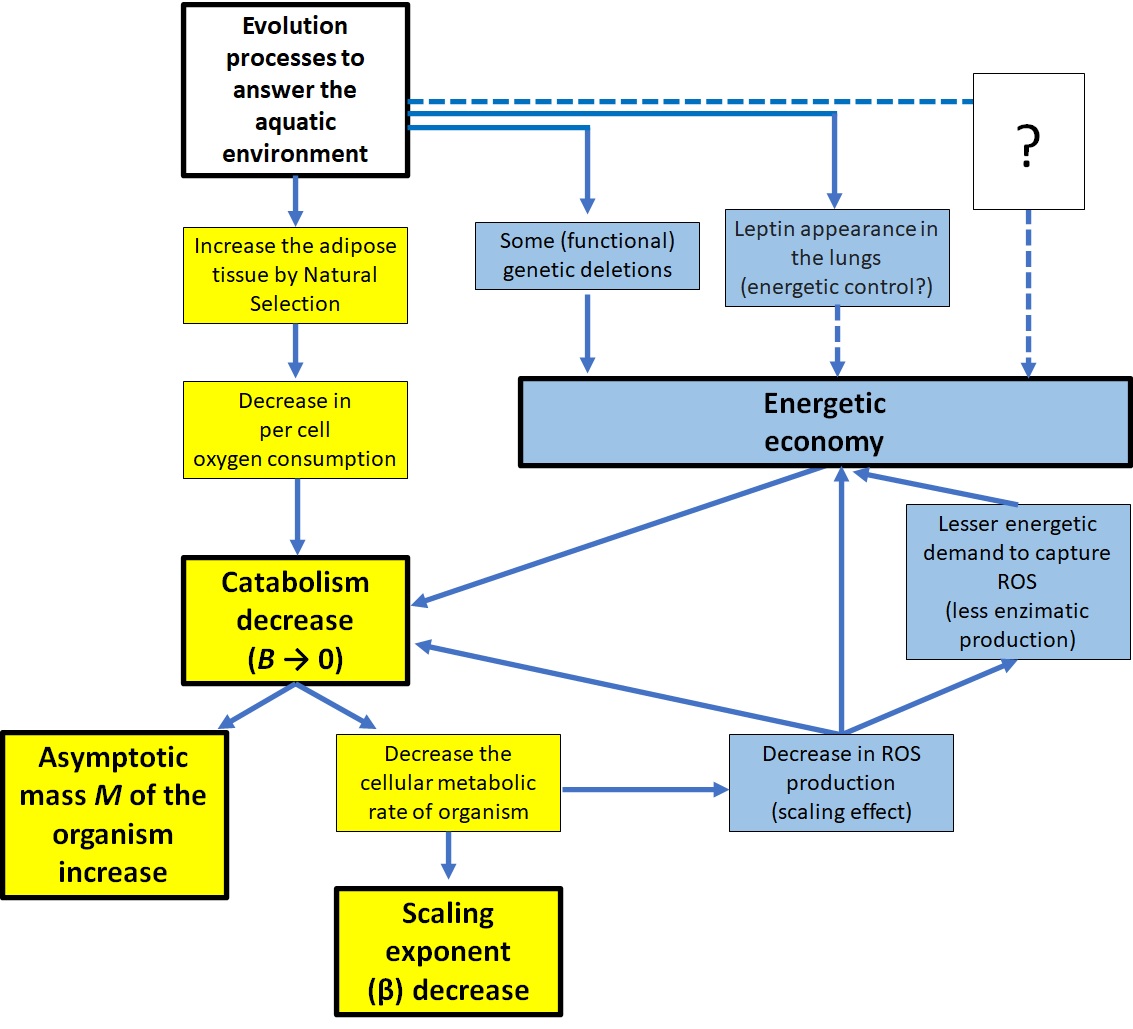}
\caption{\label{fig_diagrama2}
Diagram presenting all the events that follow the evolution process to answer the aquatic environment. The idea, basically, is that fat accumulation in marine mammals, an essential feature in this taxonomic group, triggers a sequence of energetic outputs related to scale (expressed by the metabolic parameters $\beta$, $B$ and $M$). In yellow, it is presented the events predicted by the mathematical model proposed in section~(\ref{sec_catabolism_fat}). In blue are presented some possible biological events behind the mathematical prediction, which still need to be elucidated, and the bold boxes represent the most important events. 
The accumulation of fat then leads to a decrease in the cellular metabolic rate (due to the scaling properties of adipocytes), which also implies a decrease in catabolism ($B$). Finally, as an answer to the constraint given by Eq.~(\ref{eq_M}), the organism can either decrease the scale constant ($\beta$) or increase the size ($M$), or both.}
\end{center}
\end{figure}

%In this section, we present a possible explanation to the empirical evidence described in the previous section, 
Next we describe our explanation the empirical evidence, 
especially regarding the minimisation of catabolism and the scaling exponent in marine mammals.  We argue that \textit{the large proportion of adipose tissue in these animals} can be the cause of this minimisation. 
%A mathematical model is presented in the next subsection to justify such argumentation.
We present a mathematical model to justidy this argument.
%The scheme of all processes involved in this hypothesis is presented on the diagram of Fig.~(\ref{fig_diagrama2}). 
The processes involved in this hypothesis are summarized in Fig.~(\ref{fig_diagrama2}).
A final consequence of this causal process is the increase of the asymptotic mass of these mammals.
% cause-effect

%First of all, it is essential to understand why these marine mammals have such a large proportion of adipose tissue.

%The evolutionary pressure in marine environmental forces the increase of adipose tissue in these mammals' bodies, due to some benefits of the fat blubber, as the ones already discussed in the introduction. 
The evolutionary pressure in the marine environmental forces the increase of adipose tissue in these mammals' bodies, due to benefits of the fat blubber. 
%as an aid in flotation, thermal insulation, aid in locomotion, and increase swimming efficiency by smoothing the body contour \cite{Wang2015}. 
But, most important in the context of this work is that the large proportion of adipose tissue reflects the storage and demand for energy \cite{Davis2020,Grob2020,Castrillon2017}.
%But, most important in the context of this work is that a large proportion of adipose tissue could reflect in storage and demand for energy \cite{Davis2020,Grob2020,Castrillon2017}.
Moreover, adipose tissue has a structural function \cite{Kershaw2018} and a scaling response, once that blubber thickness and body fat-mass composition scales with body mass \cite{Ryg1993}.

%Moreover, adipose tissue has structural function \cite{Kershaw2018}, and scaling response, once that blubber thickness and body fat-mass composition scales with body mass \cite{Ryg1993}.

%\sout{Some baleen whales have migratory behavior  with migration trajectories poorly described \cite{Davis2020}. For example, the humpback whale (\textit{Megaptera novaeangliae)} migrates to temperate waters for energetic acquisition \cite{Grob2020} associated with a significant period of weight loss reflecting in changes in blubber thickness, blubber weight, and body girth \cite{Castrillon2017}.} 

\begin{figure}
	\begin{center}
	        \includegraphics[width=\columnwidth]{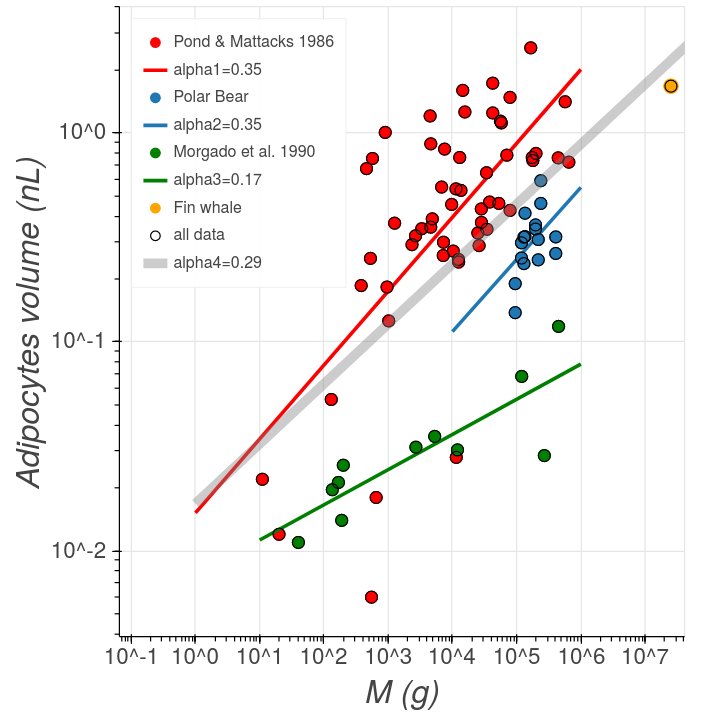}
			%\includegraphics[width=\columnwidth]{Mass-adipocyte-relationship-Pond.png}
			%\includegraphics[width=\columnwidth]{Mass-adipocyte-relationship-savage.png}
        %ncludegraphics[width=\columnwidth]{v_adipocutes.png}	
   	\caption{\label{fig_adipocytes_dara}
   	Relation between adipocyte size (volume) and organism's mass ($M$) for some taxonomy groups. It is evident the growth of adipocyte size with body mass; however, the scaling exponent value $\alpha$  is not well established. For instance, in the case of  Pond \& Mattack (1986) 's work  \cite{pond1986} and polar bear  \cite{pond1991},  the data suggest $\alpha = 0.35$ (in red and blue, respectively). However, Morgado et al.'s data suggest $0.17$ (in green). When all points are put together one has $\alpha = 0.29$ (gray line). 
The data for fin whales (\textit{Balaenoptera physalus}) is also presented (orange dot).
The data of  Pond \& Mattacks  (1985)  \cite{pond1987} is in respect to adipocytes in intra-abdominal, superficial and intermuscular depots, in species of the orders: Artiodactyla, Carnivora, Cetacea, Chiroptera, insectivora, Lagomorpha, Perissodactyla, Pinnipedia, Primates and Rodentia. In the case of polar bears (\textit{Ursus maritimus}) \cite{pond1991}, it was used adipocytes located behind the eye and around ocular muscles. In the fin whales study \cite{pond1987}, it was used the blubber and internal adipose tissue. Finally, in  Morgado 1990 \cite{Morgado1990}, the author analyses adipocytes of the skin from  species of the orders: Rodentia (3 species), Felidae (1 species), Canidae (1 species), Artiodactyla (4 species).}
%Mus musculus,Mesocricetus auratus,Mesocricetus auratus,Rattus norvegicus,Rattus norvegicus,Felis catus,Canis familiaris,Ovis aries,Sus scrofa,Equus caballus,Bos taurus.
%Note that these data  dizem respeito a especies muscalares (pouca gordura).  
  \end{center}

	\end{figure}

Adipose cells respond differently to scale in comparison to other cells.  While the mass of \textit{non-adipose (typical) cells}, say  $m_c^{(n)}$, is scale invariant \cite{Chan2010,Savage2007},  %($m_c^(n) \sim M^0$),  
some empirical studies have shown that the mass of the adipose cells, say $m_c^{(ad)}$,  increase with the organism's size $M$ by the form 
\begin{equation}
m_c^{(ad)} \sim M^{\alpha} \, ,   
\end{equation}
as suggested by the data shown in Fig.~(\ref{fig_adipocytes_dara}). 
Note that a scaling property between adipose cells volume and body mass is evident, and consequently, 
%a scaling property between adipose cells mass and body mass if we consider cellular density is invariant \cite{Savage2007}, 
adipose cells mass and body mass also scale if we consider cellular density to be invariant \cite{Savage2007},
which means $\alpha>0$. 
However the value of this scaling exponent cannot be precisely determined.
Savage et al.\ \cite{Savage2007} argue that $\alpha = 1-\beta$ in order to obey  
Kleiber's law (see supplementary material~(\ref{append_strategies})).
%Nevertheless, we will see that the value predicted by them does not explain the data presented in the previous section (more details in the next section). 
Nevertheless, we will see that the value predicted by them does not agree with the data. 

%strategy do not work for this context, because if the fat tissue obeyed  such strategy (i.e. $\alpha = 1-\beta$) the presence of fat tissue will promotes scaling economy. 

%\textcolor{red}{Diego leu ate aqui (mas nao leu of captions)}

Other important scale difference between 
adipose and non-adipose cells is concerning  the cellular metabolic rate. 
While in non-adipose cells 
their cellular metabolic rate, namely $R_c^{(n)}$,  decrease with the organism mass by the form  $R_c^{(n)} = m_c^{(n)} R_0 M^{\beta -1}$ 
(from Eq.~(\ref{eq_Kleiber's Law})), 
%given that $R_c^{(n)} = \frac{R}{N_c})$), 
the adipocytes maintain scaling invariant
its cellular metabolic rate, namely $R_c^{(ad)}$  \cite{Savage2007}, that is $R_c^{(ad)}\sim M^0$. 
These properties are organized in table~(\ref{table_cells}). In this way, the typical cells present scaling invariant mass but their cellular metabolic rate decrease with the size of the organism. 
However, the  \textit{adipocytes} does the opposite: they increase their mass with the size of the organism and maintain scale-invariant their energy demand and consequently their cellular metabolic rate.

\begin {table}
\begin{center}
	%\begin{tabular}{|p{2.5cm}|c|c|c|}
	\begin{tabular}{|c|c|c|c|}
		\hline   
		 & typical cells  & adipose cells \\
		\hline 
		cell mass & $m_c^{(n)} \sim M^0$ & $m_c^{(ad)} \sim M^{\alpha}$ \\
		\hline 
		cellular metabolic rate& $R_c^{(n)} =m_c^{(n)} R_0 M^{\beta -1}$  & $R_c^{(ad)} \sim M^0$ \\
		\hline 
	\end{tabular}
	\caption{ \label{table_cells} Scaling properties of adipose and non-adipose (typical) cells.}
\end{center}
\end{table}

%$R_c^{ad} \sim M^0$ . 
%($m_c \sim M^0$  and $R_c \sim  M^{1-\beta}$).
% with average metabolism relatively constant (and slow compared to a muscle cell or neuron, for example).

 The consequence of these properties, as will be shown by the mathematical model presented in the following subsection, is that when we have an animal with a relatively large amount of adipose tissue, as is the case of marine mammals,  the
 cellular metabolic rate decreases compared to other animals with the same mass but with proportionally smaller fat tissue. 
 Having said that, we hypothesise that the metabolic minimisation in these fat-rich animals results from the very accumulation of fat. 
 That is because fat tissue has reduced energetic demand and lower metabolism, which is expressed by smaller values of $B$ and $\beta$ values, and in the increase of the body mass $M$, conform presented in previous section. 
 This will be better explained by the mathematical model that will be presented now.

 %whole-body (global) metabolic rate per cell ($R_c = R/N$)  

%global average cellular energetic demand. 

\subsection{Mathematical Model}\label{sec_math_model}

For  an adult organism (that is $dN/dt =0$), the metabolic rate can be written (from Eq.~(\ref{eq_edoR})) as 

\begin{equation}\label{eq_R_norm_adip}
R = R^{(n)} + R^{(ad)} = N_n R_c^{(n)} +    N_{ad} R_c^{(ad)} \, ,
\end{equation}
where $R^{(n)}$, $N_n$ and $R_c^{(n)}$ (total metabolic rate, number of cells and cellular metabolic rate, respectively) are quantities related to non-adipose (typical) cells, and $R^{(ad)}$, $N_{ad}$ and $R_c^{(ad)}$ are quantities related to adipocytes.
Of course, we are considering an idealized situation that the organism is composed only by these two kinds of cells. 
If that is the case, then the total number of cells and the total mass must obey $N = N_n + N_{ad}$ and 
$M = M_n + M_{ad}$, respectively.
If $m_c^{(n)}$
and $m_c^{(ad)}$ are the cellular mass of the non-adipose and adipose cells, respectively, then $M_{n} = N_{n} \cdot m_c^{(n)}$ and 
$M_{ad} = N_{ad} \cdot m_c^{(ad)}$. 
Consequently, Eq~(\ref{eq_R_norm_adip}) yields

\begin{equation}
R = M_n \frac{R_c^{(n)}}{ m_c^{(n)}} +     
M_{ad} \frac{R_c^{(ad)}}{ m_c^{(ad)}} \, .
\end{equation}

Using the scaling properties of these cells (see table~(\ref{table_cells})), the equation above yields

\begin{equation}\label{Eq_R2}
R = M_n R_0 M^{\beta -1} + \textrm{cte} \cdot M_{ad} M^{-\alpha} \, .
\end{equation}

Lets now introduce the parameter $\lambda$ which represent the proportion of fat mass in the whole organism body. In this sense,  we consider,  by hypotheses,that 

\begin{equation}\label{hip1}
M_{ad} \equiv  \lambda M\, ,    
\end{equation}
and

\begin{equation}\label{hip2}
M_{n} \equiv (1-\lambda) M\, .     
\end{equation}
This parameter lies between $0 \le \lambda \le 1$, and these extreme cases means: $\lambda = 0$, an organism with no fat tissue; and $\lambda =1$, an (unrealistic) organism formed purely by fat tissue. A real organism must be something in between this two extreme situations.
For instance, in the case of \textit{Balaenoptera physalus}, that has around 30\% of the total body mass composed by fat tissue \cite{Pond1978}, we would have $\lambda = 0.3$.

\begin{figure}
	\begin{center}
	\includegraphics[width=\columnwidth]{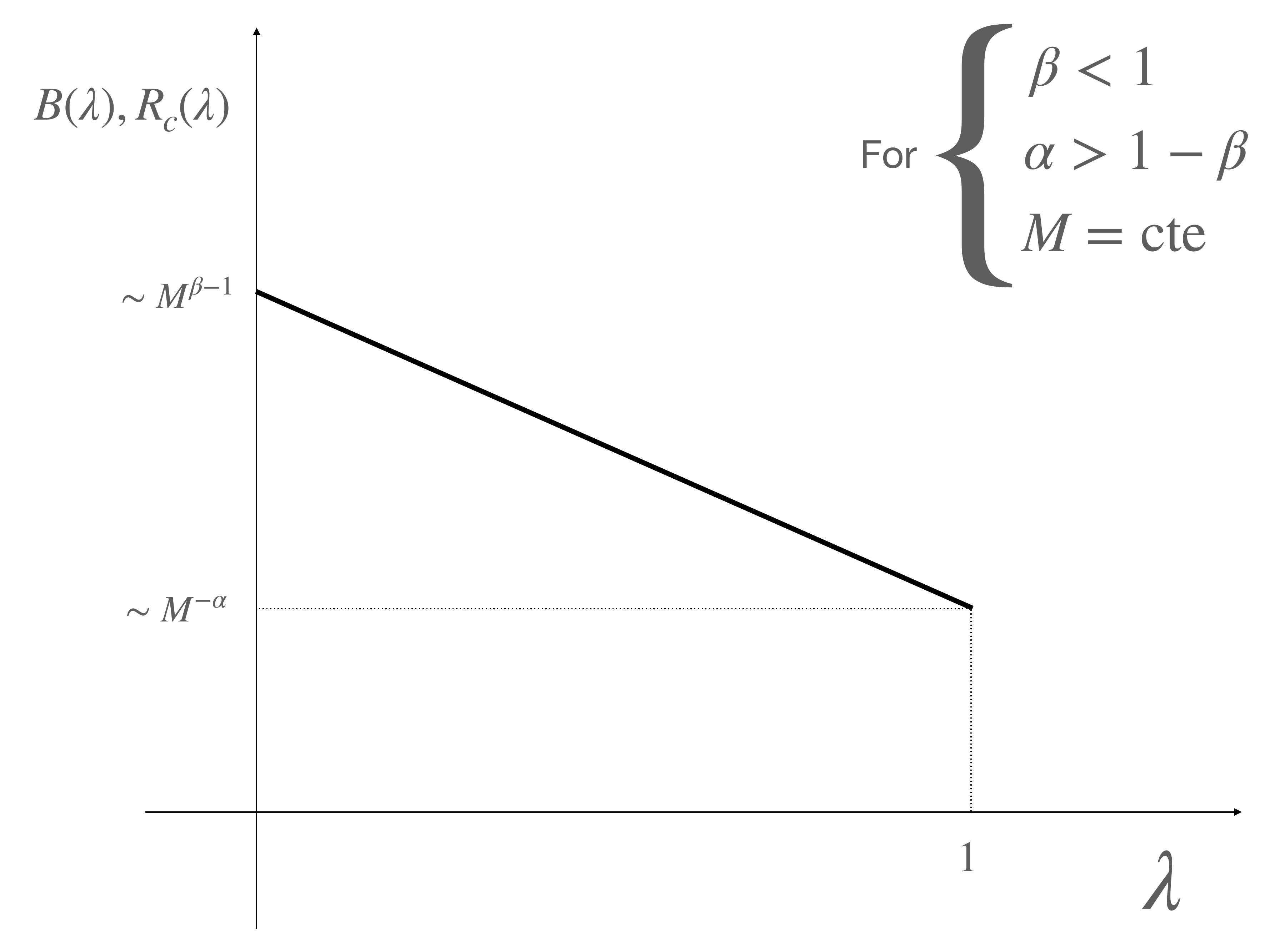}{\caption{ \label{fig_B-x-lambda} 
According to the scaling properties of the adipocyte  and the model proposed (Eq.~(\ref{Eq_Rc_lambda})), 
	 both cellular metabolic rate ($R_c$) and catabolism constant ($B$) decay with the proportion of fat tissue ($\lambda$) in the organism  
	 when $\alpha > 1- \beta$. That result suggests a scaling energetic economy for organism with larger  proportion of fat tissue.
	}}
	\end{center}
\end{figure}

Introducing such hyphotesis in Eq.~(\ref{Eq_R2}) yields

\begin{equation}\label{Eq_R_over_M}
\frac{R}{M} = (1-\lambda)R_0 M^{\beta -1} +
\lambda \cdot \textrm{cte} \cdot  M^{-\alpha} \, .
\end{equation}
In the appendix~(\ref{append_RN_RM}) we show that $R/M \sim R/N$  for $M$ sufficiently large. Also, if we identify  the organism cellular metabolic rate as $R_c = R/N$, one has from the above equation that

\begin{equation}\label{Eq_Rc_lambda}
R_c (\lambda) \sim   R_0 M^{\beta -1} -
\lambda \left( R_0 M^{\beta -1}-  \textrm{cte} \cdot  M^{-\alpha} \right) \, .
\end{equation}
That is, we write the cellular metabolic rate of the organism as a function of the proportion of fat tissue: $R_c = R_c (\lambda)$.  
Note that the total absence of adipocytes ($\lambda =0$) yields  $R_c (\lambda =0) \sim  R_0 M^{\beta -1}$, which is consistent with Kleiber's law; 
while for an organism formed only by adipocytes  ($\lambda=1$) one has $R_c (\lambda =1) \sim  M^{-\alpha}$. The result~(\ref{Eq_Rc_lambda}) also says that, when  $\alpha > 1-\beta$, 
the presence of adipose cells ($\lambda > 0$) implies that the cellular metabolic rate will decrease (scaling economy for the organism),  
as shown in Fig.~(\ref{fig_B-x-lambda}).
In other words, {\bf such result suggests that when two organisms of the same size $M$ are compared, the one with more fat tissue will have a lower average cellular metabolic rate}. That is, the presence of adipose cells promotes an organism's energetic economy.

One  point that calls attention to this result is that if the 
adipose cells follow the 
Savage et al. strategy 2 \cite{Savage2007} (see supplementary material~(\ref{append_strategies})), that using the present notation means  $\alpha = 1-\beta$, then the cellular metabolic rate will not diminish with the fat tissue proportion. 
However, of course,  the precise $\alpha$ value is inconclusive (see Fig.~(\ref{fig_adipocytes_dara})), and more careful empirical studies are needed in this direction.

%In addition, according to the values of alpha two regimes can emerge from eq when the mass is sufficiently large: i) when alpha > 1-\beta the first term on the left of the eq ??  dominates, which implies that Rc ~ M^beta-1; and ii) alpha< 1 - beta, when the secound term on the right dominates, and then Rc ~ M ^-alpha. 

\subsection{Connection with the catabolism, scaling exponent and organism size}

From Eq.~(\ref{Eq_B}) and considering only adults organism (and then the creation energy $E_c$ does not plays any rule) then $B \sim R_c$. That is, the catabolism constant is, in fact, directly related to the cellular metabolic rate, and it is valid the same properties that came with the result~(\ref{Eq_Rc_lambda}). In this way,
when $\alpha> 1-\beta$,  the catabolism  decrease with the increasing  of the proportion of adipose tissue ($\lambda$). In summary, {\bf as a hypothesis to explain the data presented in the previous section, we suggest that the bigger proportion of fat in marine mammals, caused by evolutionary pressure,  is one of the components responsible for minimising their catabolism ($B\to 0$). }
%If $B$ shows minimum values, this has a direct effect on the organism's metabolic rate. %Cetaceans can reach  55\% of blubber ($\lambda =0.55$) which, according to the idea proposed here and together with the scaling properties of the adipocytes cells, impacts directly on the organism metabolic rate.

We can also infer about the scaling exponent $\beta$ and the organism size with such results. As we saw in the previous section, any change in the ontogenetic parameters ($A$, $B$, $\beta$, and $M$) must be done constrained to the relation~(\ref{eq_M}). {\bf 
By the mathematical model proposed here, we saw that the reduction of cellular metabolic rate,  and consequently the decrease of  $B$,  is followed by i) the reduction of $\beta$, or ii) the increase of $M$, or both,  in order to maintain constrained to the relation~(\ref{eq_M}).}

For instance, Eq.~(\ref{eq_M}) can be written as 

\begin{equation}
\beta = 1- \frac{\log(A/B) }{\log{M}}\, .    
\end{equation}
Note then that if we decrease $B$ maintaining $M$ and $A$ fixed, then $\beta$ must diminish. On the other hand, Eq.~(\ref{eq_M}) can also be written as

\begin{equation}\label{Eq_Mass_catabol}
M \sim B^{\frac{1}{\beta -1}}    
\end{equation}
for $A$ fixed. Note then that by this equation, if $\beta<1$, then a decrease in $B$ implies that  $M$ increases. 
Therefore, the minimisation of $B$, maintaining $\beta$ fixed, implies automatically in the organism mass increasing (given the constraint imposed by Eq.~(\ref{eq_M})).

In conclusion, the  $B$ minimization (due to the increase in fat tissue) is automatically followed by the decrease of $\beta$  and also by the increase in the organism size $M$. % It is all a consequence of the constraint given by  Eq.~(\ref{eq_M}).
The shcheme~(\ref{fig_diagrama2}) shows the sequence of events that are chained thanks to evolutionary pressure for fat acquisition.

Alternatively, we can interpret the Eq.~(\ref{eq_M}) from the evolutionary terminology. It has well described the \textit{trade-offs} where a biological attribute has a compensatory effect on others. So, Eq.~(\ref{Eq_Mass_catabol} suggests the mandatory effect of $B$ on $M$ (from the very particular adipocyte metabolism impacting all adipose tissue) and a secondary effect of $B$ on $\beta$ (from the lower oxygen demand by adipocytes impacting in all adipose tissue). It is a phenomenon with microscopic origins, where adipocytes have their own evolutionary specialization, allowing new functional assignments on a higher scale (adipose tissue), reverberating in macroscopic attributes.

To sum up, one can conclude by the model that the accumulation of fat  leads to a decrease in the cellular metabolic rate (due to the scaling properties of adipocytes) and decreasing  in catabolism ($B$).
Finally, as an answer to the constraint given by Eq.~(\ref{eq_M}), the organism can either decrease the scale constant ($\beta$) or increase the size ($M$), or both.

%\subsection{Large Mass in aquatic mammals}

%The slower metabolism in marine mammals has unless two hypothetical consequences: i)  increase in the mass of the organism, once that the asymptotic mass depends directly on the metabolic scaling exponent and catabolism constant, as described by  Eq.~(\ref{eq_M});  and ii) reduction in per-cell oxygen consumption, due to the decrease in energy demand caused by the larger proportion of adipocytes inside the adipose tissue, mainly in blubber.

\section{Biological foundations}
\label{sec_biological}

%We show in this section how some genetic, physiological and biochemical factors cooperate to the global catabolic minimization, where it reverses to the energy economy, which can be allocated in larger body mass until the gigantism occurs. Moreover, we show how the fat tissue's scale properties impact metabolism. We defend that adipocytes, with their particular scaling properties, can affect and maintain Cetaceans healthy even if they (seems) to be obese, clarifying some intriguing paradoxes.

In this section, we propose some biological foundations to explain the metabolic-catabolic minimization and the gigantism effect in aquatic mammals.

\subsection{Scaling exponent and cell oxygen consumption}

It is crucial to speculate or think about what the minimization of $\beta$ means in biological terms. 
Darveau et al.  \cite{Darveau2002}  have shown a positive relationship between oxygen consumption and the value of the scaling exponent. It 
suggest  that a smaller value of $\beta$ in those fat-rich mammals maybe it is a consequence of the reduction of per-cell oxygen consumption.
This reduction, in turn, is caused by their metabolism reduction given by the scaling properties of the adipose cells (see schematic diagram in Fig.~(\ref{fig_diagrama2})).

\subsection{Genes related to fat accumulation in adipocytes}

Positive selection has been detected in some genes (belonging to the ACSL gene family) of cetaceans during their adaptation to the aquatic environment, which implied an association with enhanced triacylglycerols synthesis and thickened blubber \cite{Wang2015} Such genes increase the ability to capture free fatty acids and triacylglycerols in the circulation, which accelerate the increase of the adipocytes size. In this way, the adipocytes size expansion is directly linked to the amount of fat accumulated in these cells. 
Moreover, some specific genes related to fat synthesis and production can be found in a common ancestor of fat-accumulating marine mammals, indicating a strong adaptive selection in such genes \cite{Wang2015}. Since blubber composition is made mainly by adypocites, and Cetacean health can be diagnosed by blubbler thickness,\cite{Derous2020}, such genes could be crucials to these animals.

\subsection{Leptin and adipocyte}

In terrestrial mammals, adipose tissue acquired the property of secreting hormones; leptin, a hormone produced and secreted by adipocytes, is strongly related to adipocyte hypertrophy, as seen in rats and humans \cite{Maffei1995}. Interestingly, in addition to finding significant leptin gene expression in the blubber of seals, an unusual and significant expression of leptin was also observed in their lungs \cite{Hammond2005}.
These findings suggest that maybe leptin promotes the energy balance in the lungs, and not only on the adipose tissue.
In the previous section, we saw that the decrease in the oxygen consumption could be associated with the metabolism  minimization.
Currently, there is evidence pointing to adaptations where leptin started to participate in the respiratory processes of marine mammals, including whales, mainly in conditions of lower oxygen supply (hypoxia) \cite{Yu2011}.
To sum up, the leptin participation in respiratory control suggests that the respiratory system also contributes to minimising metabolism, but the literature still lacks experimental evidence and mechanistic explanations.

\subsection{Fishes and muscles}

Usually, fishes have 50-60\% of their biomass composition of muscle, and this trend is contrary to marine mammals. They are ectothermic, and their metabolism is strongly influenced by environmental variations of the temperature, with a positive relationship between temperature and body mass \cite{Audzijonyte2020}. 
Moreover, some tilapia strains are also selected to be thermo-resistant to environmental variations \cite{DosSantos2013}. Since most of the data from tilapia used in this study come from zootechnical studies -  that is, in optimal growth environments (with optimal temperature) - we were interested to see some response to this artificial scenario in our numerical results. However, our analysis (not shown in this study) did not indicate any effect of these artificial improvements in $\beta$ values, but yes in the parameters $A$ and $B$,  and consequently in $M$ (from Eq.~(\ref{eq_M})). It suggests that $\beta$ has solid genetic and evolutionary determinants, but is not susceptible to environmental pressure.

The linear trend between $A$ and $B$ (Fig. (\ref{fig_BxA})) confirms that fast-developing animals (and shorter lifespan) are directly affected by both parameters. In fishes, a higher anabolism constant could be related to higher biomass muscle synthesis with faster maturity age. On the other hand, a higher catabolism constant could be related to a higher energetic cost to muscle maintenance. Muscle tissue has biological properties very different from adipose tissue, with higher mitochondrial concentration (reflects the higher metabolic activity) and in fishes, where they employ almost all muscle body mass to contract and to swim, this higher energetic demand to maintain muscle function is proportionally higher. So these zootechnical improvements in tilapia seem to respond mainly in $A$, reflecting in $B$ (given  Eq. (\ref{eq_M}, see Fig. (\ref{fig_BxA}) with more conservative values to $\beta$.

%\textcolor{blue}{Agora entendi a equacao 8. Com esse exemplo da tilapia, isso sugere que, de fato, a equacao 8 e um medidor de trade-off. Veja o paragrafo que inseri no final da sessao 4 para explicar esse vinculo com argumentos evolucionarios. Trade-off pertence a terminologia de estudos sobre evolucao e a Teoria das Historias de Vida. Entao, a equacao 8 mostra como os parametros metabolicos se equilibraram entre si. Agora, isso e muito novo, porque esse equilibrio e funcao de $\beta$. Se beta e bastante conservativo, modificando apenas com shits evolutivos bruscos, isso sogere que esses shifts levam a implicacoes metabolicas muito impactantes, remetendo o paper de Delong et al. Nosso paper descreve mais 2 shifts, o shift que ocorreu nos Cetaceos decorrente da alta adiposidade e o shift que ocorreu nos peixes, decorrente da alta proporcao de tecido muscular. Portanto, sem querer, descobrimos 2 shifts metabolicos dentro dos organismos Metazoarios apos eles vascularizarem com expoente de 3/4 e agora estamos descrevendo uma metodologia poderosa para descrever mais shifts!!!!!!!!!}.

\section{Conclusion}
\label{sec_final}

We are conscious that the animal's size is a multiplicative effect of a series of factors. However, we offer in this work some empirical finds and a theoretical approach to argue that fat accumulation in aquatic mammals triggers a series of events that increase the size of these animals.

The pieces of evidence shown and presented here suggest that when we have an animal with a relatively large amount of adipose tissue, as is the case of marine mammals,  the cellular metabolic rate decreases compared to other animals with the same mass but with proportionally smaller fat tissue. That is due to the different scaling properties of the adipose cell in comparison to the non-adipose cells. It implies that 
fat tissue has reduced energetic demand and lower metabolism,
which is expressed by smaller values of $B$ and $\beta$ values, and in the increase of the body mass $M$. That is the case for marine mammals. 
It suggests that the metabolic minimization in these fat-rich animals results from the very accumulation of fat. 
As a consequence of these cause-effect mechanisms and given the constraint imposed by the relation~(\ref{eq_M}), we have the emergency of those big animals. 

In conclusion, one can say that the large proportion of fat tissue in marine mammals has the following hypothetical consequences: i) slower metabolism (characterized by smaller values of the catabolism constant $B$ ;  
ii) reduced scaling exponent $\beta$, which may be related to the reduction in per-cell oxygen consumption; and iii) increase in the organism's mass $M$.

\acknowledgments{
We would like to thank Diego Rybski and Vinicius Netto for the comments and precious suggestions for the first versions of this paper.
%, and Josiane O. Pinto Ribeiro for figure~(\ref{fig_animais}).
F.L.R wants to thank the financial support from the Brazilian agencies 
CAPES (process number: 88881.119533/2016-01), 
FAPEMIG  (process number: APQ-00829-21),
and 
CNPq (process number: 403139/2021-0). }

%This work was only made possible by the invaluable help of Victor Cabral and Erika Aparecida Costa Lomeu with the figures presented. 
%anonymous referees for their valuable contributions.  

\newpage

\section{References}\label{references}

%\bibliographystyle{ieeetr}

%\bibliography{william.bib, fabiano.bib}

\newpage

.  

\newpage

\appendix

\section{Ontogenetic Models}\label{app_ontogenetic}

It is presented in this supplementary material section the ontogenetic models that were used to describe the temporal dynamics of mass, length and energy of the organisms analysed, and consequently the ontogenetic parameters used. 
It is discussed here the  \textit{Bertalanffy-Richards model}, which describes the time evolution of the organism mass;  the \textit{length-weight relationship}, which is the  empirical evidence that the organism's mass and length  obey a power law relation; 
version of the Bertalanffy-Richards Model for the time evolution of length; and finally, the \textit{West-Banavar model}, which describes the time evolution of the organisms' maintenance energy.
 
%\textcolor{blue}{Fabiano, voce esta chamando a equacao A15 de Equacao de West et al. No entanto, West criou o paradigma de $\beta$ = 3/4 e isso deixou um rastro negativo. Eu propus de chamar de Equacao de West-Banavar porque Banavar e seu grupo tiveram coragem para criticar West et al, colocando alguns argumentos para criticar a ideia de universalidade por detras dessa equacao, incluindo a insercao de um funcao dependente da massa assintotica, que reforca o quao importante e analisar esses fenomenos apos dm/dt=0. Verificamos aqui que certos fenomenos metabolicos so sao plenamente manifestados apos o ponto de infexao, ou biologicamente, quando o organismo esta atingindo a maturidade. Gostaria q voce lesse o paper de Banavar et al para saber se vale a pena a gente evocar esse grande fisico, ele e o grupo dele, e dar o devido valor para eles.}

\subsection{Bertalanffy-Richards Model}

%von Bertalanffy, L
% F. J. Richards, J. Exp. Bot. 10, 290 (1959).

The Bertalanffy-Richards Model \cite{bertalanffy1957,richards1959}, which was presented in the main part of the paper
(see Eq.~(\ref{edo_m})) 
describes the time evolution of the organism's  mass \cite{Cabella2011,ribeiro_tumor2017,Cabella2012a}.
The solution of this model if given by Eq.~(\ref{solucao_edo}), and it is written  in terms of the ontogenetic 
parameters  $A,B,m_0,$ and $\beta$, whose biological means was discussed in the main text.

Given that the anabolism  and catabolism constants,  $A$ and $B$ respectivelly,  are related one each other by the constraint given by Eq.~(\ref{eq_M}), it is possible to eliminate one of this parameters in the solution 
if we have available the organism's saturation mass $M$.
In this way it is possible
to write the solution of this model  in at least two way, depending on the type of parameter to be investigated. 
The first one is writing the solution in an explicit form of the catabolism constant $B$, and then eliminating the depence with the anabolism constant $A$. That is, inserting the Eq.~(\ref{eq_M}) to the solution~(\ref{solucao_edo}), which yields

\begin{equation}\label{Eq_BR-B}
m(t)=M\left[1+\left(\left(\frac{m_0}{M}\right)^{1-\beta}-1\right)e^{B(\beta -1)t}\right]^\frac{1}{1-\beta}\, .
\end{equation} 
In this case we can say that  $m(t) = m(t| M,B,m_0,\beta)$, expliciting the depentence of the solution to these four ontogenetic parameters.

Alternatively, mantaining explicit the anabolism constant by eliminating the catabolism constat, one gets

\begin{equation}\label{Eq_BR-A}
m(t)=M\left[1+ \left(\left(\frac{m_0}{M}\right)^{1-\beta}-1\right) 
e^{\frac{A (\beta -1)}{M^{1-\beta}} t}
\right]^\frac{1}{1-\beta} \, .
\end{equation} 
In this case one can write that  $m(t) = m(t| M,A,m_0,\beta)$.
Of course, these two ways of writing the model's solution only works when the saturation mass of the organism is available, which happens when $\beta<1$.

\subsection{Length-weight relationship}

The next model is in fact the empirical evidence of the  power-law relation between mass and length of organisms: the so called \textit{length-weight relationship} LWR \cite{Froese2006}. 
That is, the mass $m$ of the organism scales with its linear body length
$l$ as

\begin{equation}\label{Eq_MC}
m = a l^b,
\end{equation}
where $a$ and $b$  are parameters of this model which can be obtained by data-fitting.
The relation~(\ref{Eq_MC})  is 
commonly and traditionally described in fisheries experimental data \cite{Froese2006}, and nowadays it is applied for many purposes, for example, to predict biomass in large marine mammals \cite{Vikingsson1988}\cite{Gunnlaugsson2020}\cite{Amaral2010}\cite{Agbayani2020}, due to difficulties in experimental measurements.

The empirical parameters $a$ and $b$ carry biological information, that could be any biotic \cite{Chacon_et_al_1992a}\cite{Han2017}\cite{Crispel2013} or abiotic effect \cite{MacDonald_Lisac_1996}\cite{Kolding2008}, or a meaningful taxonomic information \cite{Frasier2015}\cite{Yi2017}.
%The LWR was valuable to revels the patterns found in this work.

%In fact, we use these facts to estimate better the numeric values of the metabolic parameters (see next section, where we describe the metodology used to estimated the ontogenetic parameters).

\subsection{Length time evolution}

The next model is the junction of LWR and
the Bertalanffy-Richards model, wich results in the the length time evolution,  which also also obey a Betalanffy-Richards dynamics.

To demonstrate this consider firstly the derivative of the LWR~(\ref{Eq_MC}), that is  

\begin{equation}\label{Eq_Lm1}
\frac{d m}{dt} = a b l^{b-1}\frac{dl}{dt} \, .
\end{equation}
Parallel  we can insert~(\ref{Eq_MC}) into~(\ref{edo_m}) to get  

\begin{equation}\label{Eq_Lm2}
\frac{dm}{dt} = A(al)^{\beta } - B a l^b \, .
\end{equation}
Equaling these two last equations, one gets the ODE

\begin{equation}\label{eq_model_L}
\frac{dl}{dt} = C l^{\delta} - D l  \, ,
\end{equation}
which shows that the length $l$ is also governed by a Betertalanffy-Richards dynamics, where 
 
\begin{equation}\label{EL1}
 C \equiv \frac{Aa^{\beta -1}}{b},
\end{equation}

\begin{equation}\label{EL2}
D \equiv \frac{B}{b} \, ,
\end{equation}
and 

\begin{equation}\label{EL3}
\delta \equiv b(\beta-1)+1 \, .
\end{equation}
In fact, Essington et al.~\cite{Essington2001} present some empirical evidences that length $l$ 
could be modelled by the Bertalanffy-Richards model.
We coined \textit{Essington et al. relationships} these three relationships (\ref{EL1}, \ref{EL2}, and \ref{EL3}), where the parameters related to length can be explained in metabolic terms.

The solution of the ODE~(\ref{eq_model_L})  has the same logistic-like behavior that emerges during ontogenetic body mass growth, that is

\begin{equation}\label{eq_comprimento}
l(t)=L\left[1+\left(\left(\frac{l_0}{L}\right)^{(1-\delta)}-1\right)e^{D(\delta -1)t}\right]^\frac{1}{1-\delta},
\end{equation}
with $L \equiv l(t\to \infty)$  been the saturation/assimptotic length, given by

\begin{equation}
L = \left( \frac{C}{D} \right)^\frac{1}{1- \delta}
\end{equation}

or symply

\begin{equation}
L  = \left( \frac{M}{a}\right)^{\frac{1}{b}}.
\end{equation}
The condition for the saturation to happen is that
$0<\delta<1$.

\subsection{Energy allocated to mantainence}

The last model to be presented is based on the seminal work of 
West et al. \cite{West2001}, which describe the temporal evolution of the energy maintenance of the organism (see also
\cite{Guiot2003a,Savage2013a,
ribeiro_tumor2017}). 
Using the idea that the energy in a organism ($R$) is used for manutence ($N R_c$) and growth $E_c \frac{dN}{dt}$, conform described by  Eq.~(\ref{eq_edoR}), then one can show that the ration between manutence energy and total energy, say

\begin{equation}
 r = \frac{\textrm{ maintenance energy} }{\textrm{total energy}} \, ,  
\end{equation}
  can be writen  (using the definition presented in section~(\ref{sec_model}))  
as
\begin{equation}\label{Eq_r1}
r(t)=\left(\frac{m(t)}{M}\right)^{1-\beta} \, .
\end{equation} 
Then, comparing this result with Eq.~(\ref{Eq_BR-A}), one gets

\begin{equation}\label{Eq_r2}
r(t)=  1+\left(\left(\frac{m_0}{M}\right)^{1-\beta}  -1 \right)e^{\frac{A(\beta -1)}{M^{1-\beta} }t} \, ,
\end{equation} 
The quantity $r(t)$   revels  universal properties, been approximately the same for a huge amount of species, as demonstrated in the original paper and in other works \cite{
Guiot2003a,Savage2013a,
ribeiro_tumor2017}.
We use this properties of universality in the present work as a additional information to calibrate the methods and to get a better estimated value for the ontogenetic parameters. More detail in the next section.

The $\beta$ exponent in~(\ref{Eq_r2})
was relaxed by \textit{Banavar et al} \cite{Banavar2002}, where it gave rise to doubts in the universality ideas suggested by \textit{West et al} \cite{West2001}. Moreover, they suggested a function acting in biomass saturation. Because this improvements helped us to make more accurate fit-to-data,  we coined  equation~(\ref{Eq_r2}) as West-Banavar (W-B) model.

%\textcolor{blue}
%{BANAVAR et al foram os primeiros questionar o conceito de universalide e o valor universal de 1/4 depois da publicacao do paper de WEST et al, sugerindo que a existe uma funcao que atua na massa maxima, que e dependente de $\beta$ e altera a curva universal. Apesar de ser um paper com poucas citacoes, foi publicado na Nature logo apos o paper de West. Banavar e fisico e atua fortemente no ramo, por isso acho justo chamar de WEST-BANAVAR}

%where $\beta$ was relaxed by \textcolor{blue}{BANAVAR et al. (2002)}, we can coin the West-Banavar (W-B) equation and to this elegant transformation we can coin this the West et al Plot. Moreover \textcolor{blue}{WEST et al. (2001)} designed $r(t)$ to revels a new biological variable, the energy allocated to growth (\textcolor{red}{Na verdade eu acho que eh a proporcao de energia alocada para manutencao, checar isso}). 

%\newpage

\section{Getting metabolic parametres with the fitting curve with data and the models}
\label{append_metod}

This supplementary material section presents the methodology used to determine the ontogenetic parameters' values from the data. 
 Those data are available on a  sheet
\attachfile{ontogenetic-trajectories-database.xlsx} 
that contain  mass ($m$), length ($l$) and time ($t$) for  43 ontogenetic trajectories.
The data were extracted from the literature and can be  classified into four groups:

\begin{itemize}

\item{
Terrestrial small-size mammals:
Yangtze vole (\textit{Microtus fortis calamorum})\cite{Han2011}, 
Sprague-Dawley rat (\textit{Rattus rattus})\cite{Crispel2013}, 
Taiwan field mouse (\textit{Apodemus semotus})\cite{Lin1993},
 Cabrera vole (\textit{Microtus cabrerae})\cite{Fernandez-Salvador2001}.}

\item{
Terrestrial medium-size mammals: anglo nubian goat (\textit{Ovies aries})\cite{Oliveira2017},
wild boar (\textit{Sus scrofa})\cite{Jezek2011}.}

\item{
Marine large-size mammals:
gray whale (\textit{Eschrichtius robustus})\cite{Agbayani2020}, 
southern elephant seal (\textit{Mirounga leonina})\cite{Boyd_et_al_1994}, 
Amazonian manatee (\textit{Trichechus inunguis})\cite{Colares_2002}\cite{Mendoza2019}, 
Antillean manatee (\textit{Trichechus manatus manatus})\cite{Colares_2002}.}

\item{
Fishes:
rainbow trout (\textit{Oncorhynchus mykiss}\cite{MacDonald_Lisac_1996}), 
Nile tilapia (\textit{Oreochromis niloticus})\cite{Chacon_et_al_1992a}\cite{SilvaJ.W.B.MachadoJ.R.NobreM.I.daS.Bezerra1992}\cite{Silva_et_al_1992a}\cite{Medri2000}\cite{Kolding2008}.}

\end{itemize}

One example of such available data is the one presented in table~(\ref{table_wale}), which shows the time evolution of mass and length in the gray whale (\textit{Eschrichtius robustus}). The methodology that will be described here compares those data  with the  ontogenetic growth equations  
(\ref{Eq_BR-B}), (\ref{Eq_BR-A}), (\ref{Eq_MC}),  (\ref{eq_comprimento}), and (\ref{Eq_r2}).
The intention is to extract, for each species, the ontogenetic parameters' values that best describe, simultaneously,  all these ontogenetic growth equations.

\begin{table}[]

\begin{center}
	%\begin{tabular}{|p{2.5cm}|c|c|c|}
	\begin{tabular}{|c|c|c|c|}
		\hline   
		Age & Age & Length & Weight  \\
		 (years)& (days)	& (cm)& (g)\\
		\hline  
0 & - & -	& - \\
\hline  
1 &	365 &	851	& 6019000\\
\hline  
2 &	730 &	922	& 7618000\\
\hline  
3 &	1095 &	981 &	9155000\\
\hline  
4 &	1460 &	1031 &	10590000\\
\hline  
5 &	1825 &	1072 &	11901000\\
\hline  
6 &	2190 &	1107 &	13077000\\
\hline  
7 &	2555 &	1136 &	14119000\\
\hline  
8 &	2920 &	1160 &	15033000\\
\hline  
9 &	3285 &	1181 &	15828000\\
\hline  
10 &	3650 &	1198 &	16515000\\
\hline  
11 &	4015 &	1212 &	17105000\\
\hline  
12 &	4380 &	1224 &	17611000\\
\hline  
13 &	4745 &	1234 &	18042000\\
\hline  
14 &	5110 &	1243 &	18408000\\
\hline  
15 & 	5475 &	1250 &	18719000\\
\hline  
16 &	5840 &	1256 &	18982000\\
\hline  
17 &	6205 &	1261 &	19205000\\
\hline  
18 &	6570 &	1265 &	19393000\\
\hline  
19 &	6935 &	1269 &	19551000\\
\hline  
20 &	7300 &	1271 &	19685000\\
\hline  
25 &	9125 &	1280 &	20094000\\
\hline  
30 &	10950 &	1284 &	20266000\\
\hline  
35 &	12775 &	1286 &	20338000\\
\hline  
40 &	14600 &	1286 &	20368000\\
 	\hline 
	\end{tabular}
	\caption{ 
    \label{table_wale} Data of mass, length and time for gray whale (\textit{Eschrichtius robustus}) got from \cite{Agbayani2020}. Similar data for others 42 ontogenetic trajectories are also available for our analyses.
    %(see ontogenetic-trajectories-database.xlsx file). %In the sheet present in the next section of supplementary material there data for all species analysed and also the references we get those data.}
    }
\end{center}
%    \centering
%    \includegraphics[scale=0.355]{Tabela-Baleia-cinza.pdf}	
\end{table}

Some ontogenetic parameters are obtained by direct curve-fitting;  others, however, can be obtained (or estimated) directly from the data, as is the case  of parameters that govern the initial dynamics, as $m_0$ and $l_0$,   and the ones that govern the saturation, as $M$ and $L$. We can use the empirical values of these parameters as a constraint to estimate the other ontogenetic parameters values with more precision.

\begin{figure*}
	\begin{center}
	\includegraphics[scale=0.355]{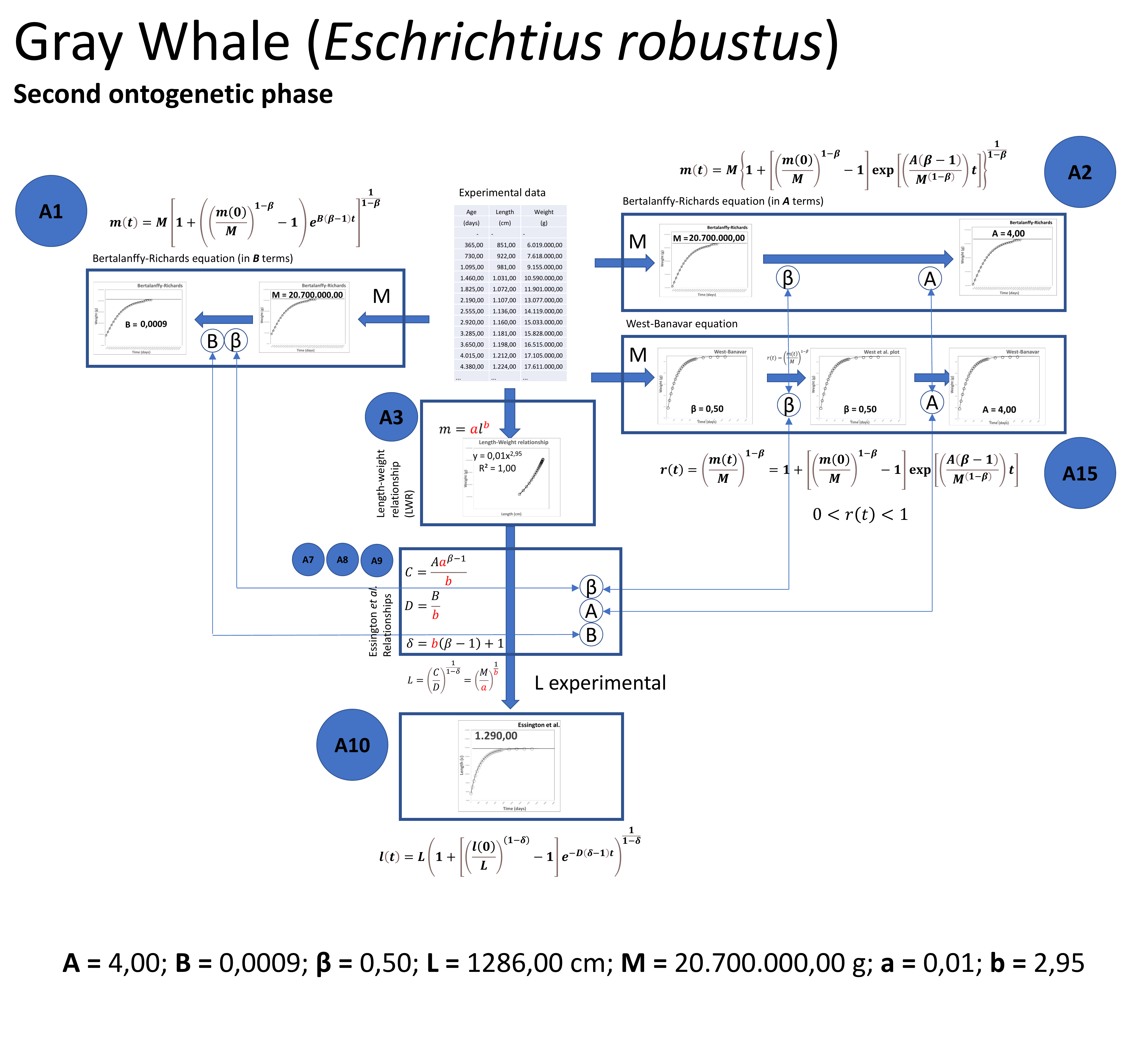}
		\end{center}
	\caption{ \label{fig_algoti}
Diagram showing how we proceed with the data and models to get the values of the ontogenetic parameters ($A$, $B$, $\beta$, $M$, $L$, $a$, and $b$)  for a specific species. The gray whale data was used in this particular example. The tick arrows show how the data are inserted into the models, while the thin arrows show how the ontogenetic parameters' values are recurrently obtained and feedback the other models. This process is fulfilled in all 42 ontogenetic trajectories available in the database.
	}
\end{figure*}

The schema drawn in  Fig.~(\ref{fig_algoti}) describes the methodology used to fit the data simultaneously with all the five ontogenetic growth curves. 
The idea is to adjust the values of the parameters interactively with all growth curves in a recurrently way. This is done step by step until the values converge.
The flowchart presented in this figure shows the order in which the growth curves are used, and the order in which the parameters are updated. 
%A sheet with all ontogenetic parameters value estimated by this methodology is presented at the attached file \textcolor{red}{(colocar link csv file)}.

It is important to stress that the studies presented in the main text of this work are built by the knowledge of only four ontogenetic parameters: $A$, $B$, $M$  and $\beta$. To support the hypotheses proposed, it is crucial to have good accuracy in these parameters values for all analysed species. That is why this methodology uses five  growth equations (Eqs.~ 
(\ref{Eq_BR-B}), (\ref{Eq_BR-A}), (\ref{Eq_MC}),  (\ref{eq_comprimento}), and (\ref{Eq_r2})) and, consequently, other ontogenetic parameters.

For instance,  only the adjust (fit) of the model~(\ref{solucao_edo}) with mass and time  data  is enough to the get  what we need (i.e.
 $A$,$B$, $M$  and $\beta$). 
However, if we restrict to this model's data analysis only, the parameters' values obtained will be very noisy and imprecise. 
 On the other hand,  using this fit  together and constrained to other growth equations
 allows us to get better and more precise values for the parameters (for instance, using $a$, $b$, and $L$ from the mass-length relation).
 %four models besides just one and including the data of length growth incorporates more information that 
In fact, the final values obtained have an $r^2$ always higher than   $0.87$ for all analysed species.

\section{Stategies of Savage et al. }\label{append_strategies}

The Kleiber's Law ~(\ref{eq_Kleiber's Law}) and the relations between organism and cell properties (given by: $R = N R_c$ and $M=N m_c$) implies that the ratio between cellular metabolic rate $R_c$ and cellular mass $m_c$ must obey the scaling

\begin{equation}
\frac{R_c}{m_c} = R_0 M^{\beta -1 }  \, .  
\end{equation}
Based on this,  Savage et al.\cite{Savage2007} namely two possible scaling scenarios the cell must obey. They are:

\begin{itemize}
    \item {\bf strategy 1}: cellular average mass remains fixed - scaling invariant -, but the cellular metabolic rate must be scaling dependent. 
    That is 
    
    \begin{equation}
      m_c \sim  M^0  \text{  and  }  R_c \sim M^{ \beta -1} \, .  
    \end{equation}
    This strategy is followed by typical cells;

\item {\bf strategy 2}:  cellular metabolic rate remains fixed, but then cellular mass must be scaling dependent. That is 

  \begin{equation}
  m_c \sim  M^{1 -\beta} \text{ and } 
  R_c \sim M^0 \,  .
   \end{equation}

%This strategy is followed by the adipocytes cells 

\end{itemize}

These strategies reflect the fact that cellular metabolic rate and cellular mass cannot both remain scale-invariant simultaneously.

\section{Relation between $R_c = R/N$  and $R/M$}\label{append_RN_RM}

In this supplementary material section we will show that when we have organism formed by two types of cell: typical and adipose cells, then  
 $R/N \sim R/M$  when $M$ is sufficiently large.
Identify such relation is important because in the mathematical model presented in section~(\ref{sec_math_model}) we gets $R/M$, however the goal is to find a expression for $R_c = R/N$. 
If we were considering an organism composed only of typical cells, the relationship between $R/M$ and $R/N$ is direct, but this is not exactly the case when we have an organism with a significant amount of 
adipocytes, which have different scaling properties.

First of all note that the total number of cells (the sum of the number of typical cells and the number of adipose cells) can be write as

\begin{equation}
N = N^{(n)} + N^{(ad)} = \frac{M_n}{m_c^{(n)}} +     \frac{M_{ad}}{m_c^{(ad)} }  \, .
\end{equation}
And from the hypotheses~(\ref{hip1}) and~(\ref{hip2}), 

\begin{equation}
N \sim  (1-\lambda) M + \lambda \cdot \textrm{cte} M^{1-\alpha} \, ,  
\end{equation}
which give us how $N$ scales with the 
organism mass when it is composed by a  proportion $\lambda$ of fat tissue.
Note that for $\lambda = 0$ (organism composed only by typical cells),  simply $N \sim M$.  However for any fixed $\lambda$ one has that    

\begin{equation}
N \sim  M + \textrm{cte} \cdot M^{1-\alpha} \, ,  \end{equation}
But note that if $\alpha > 0$ (which is supported by the empirical data, see Fig.~(\ref{fig_adipocytes_dara})), then the second tern on the right of this equation above is always smaller than the first. Then, for $M$ sufficiently large only the first term on the right of this equation plays the rule, that is $N \sim M$ for $M$ sufficiently large and for individuals composed by both typical and adipose cells. In this way, 
\begin{equation}
R_c = \frac{R}{N} \sim  \frac{R}{M}  \, ,
    \end{equation}
as we wanted to demonstrate.

\section{Data Sheets}\label{append_planilhas}

In this supplementary material section is attached all the data (and the respective references) used in the present work.

In the file 
\attachfile{ontogenetic-trajectories-database.xlsx}
it is presented all ontogenic trajectories used. 
That includes mass, length, and age for 42 species.

In the file 
\attachfile{relacao-massa-corporea-volume-adipocito.xlsx}
It presents all the data related to adipocytes used in this work. It includes adipocyte volume,  body mass, and the references in which the data was taken.

\end{document}